\documentclass[a4paper,11pt]{article}
\pdfoutput=1
\usepackage{jheppub}

\newcommand{\mathsym}[1]{{}}

\newcommand{\ba}{\begin{array}}
\newcommand{\ea}{\end{array}}
\newcommand{\bal}{\begin{align}}
\newcommand{\eal}{\end{align}}
\newcommand{\be}{\begin{equation}}
\newcommand{\ee}{\end{equation}}
\newcommand{\beqa}{\begin{eqnarray}}
\newcommand{\eeqa}{\end{eqnarray}}

\def\321{$SU(3)\times SU(2)\times U(1)$}

\def\ca{\cos\alpha}
\def\cb{\cos\beta}
\def\sa{\sin\alpha}
\def\sb{\sin\beta}

\title{Leptonic Precision Test of Leptophilic Two-Higgs-Doublet Model}

\author{Eung Jin Chun}
\author[1]{and Jinsu Kim\note{Corresponding author.}}
\affiliation{Korea Institute for Advanced Study, Seoul 02455, Korea}

\emailAdd{ejchun@kias.re.kr}
\emailAdd{kimjinsu@kias.re.kr}

\abstract{
The type X (lepton-specific) two-Higgs-doublet model at large $\tan\beta$  becomes leptophilic 
and thus allows a light pseudoscalar $A$ accommodating the observed 
muon $g-2$ deviation without conflicting with various hadronic constraints.
On the other hand, it is strongly constrained by leptonic precision observables such as 
lepton universality test in the neutral and charged currents.  Treating all the lepton universality data in a 
consistent way, we show how the current data constrain the parameter space of $m_A$ and  $\tan\beta$ 
for given degenerate masses of heavy Higgs bosons $H$ and $H^\pm$. 
While no overlapping region is found at $1\sigma$, a sizeable region is still viable at $2\sigma$ 
for $H/H^\pm$ masses at around 200$\sim$400 GeV.
}

\begin{document}
\maketitle
\flushbottom

\section{Introduction}

The electroweak symmetry breaking may involve more than one  Englert-Brout-Higgs (EBH) boson
beyond the minimality of the original Standard Model (SM).  Such a two-Higgs-Doublet Model (2HDM) 
suffers from generic flavor violation which can be naturally  resolved by a (softly broken) 
$Z_2$ symmetry enforcing a EBH doublet couple to one type of fermions. 
Depending on how a EBH doublet couple to up/down-type quarks and charged leptons, there appear 
four types of 2HDMs which have been studied extensively \cite{2hdms}.  

Since the first measurement of the muon anomalous magnetic moment, $a_\mu = (g-2)_\mu/2$,  
by the Muon $g-2$ Collaboration in E821 at BNL \cite{bnl0102},
various progresses have been made in experimental inputs as well as  theoretical calculations to reduce the uncertainty by a factor of two or so and thus establish a concrete $3 \sigma$ discrepancy \cite{amu}.
%
After the BNL announcement, many studies have been made in the context of 2HDM Type II \cite{type2,maria0208} and X \cite{cao0909}. It has been known that the discrepancy in the muon $g-2$ 
can be accommodated through Barr-Zee two-loop corrections \cite{bz} if the extended Higgs sector allows a relatively light pseudoscalar with large Yukawa couplings to fermions.
Recently, it was found that only the lepton-specific (or type X)  2HDM remains viable after considering 
all the updated experimental constraints \cite{broggio1409,wang1412,abe1504,chun1507,chun1511,crivellin15}.

The extra CP-even/odd Higgs bosons in Type-X 2HDM have the quark (lepton) Yukawa couplings suppressed (enhanced) by $\tan\beta$. That is, they become hadrophobic (or leptophilic) in the large $\tan\beta$ limit favored by the muon $g-2$ explanation and thus easily evade all the constraints coming from 
hadronic observables like $b\to s \gamma$ and $b \to s \mu^+ \mu^-$. We call this leptophilic 2HDM (L2HDM). However, it was noted in \cite{abe1504} that precision determination of lepton universality 
in the neutral and charged currents, i.e., in the leptonic $Z$ decays and leptonic/semi-hadronic $\tau$ decays,
could play an important role to constrain the L2HDM parameter space.

Given the importance of the lepton universality test limiting new physics effect in general, 
we re-evaluate the extra Higgs boson's contribution to $Z\to \tau\tau$ and $\tau$ decays
which have been analyzed previously in \cite{denner91} and \cite{cao0909,abe1504}.  Then, we improve
the constraints on L2HDM treating properly the full leptonic precision data on lepton universality.
The lepton universality in $Z$ decays was tested by measuring the ratios $\Gamma(Z\to\mu\mu)/\Gamma(Z\to e e )$ and $\Gamma(Z\to\tau\tau)/\Gamma(Z\to e e )$ \cite{lepew05}. Another test was carried out by HFAG in the processes of $l\to l' \nu \nu'$, $\tau \to \nu \pi/K$ and $\pi/K\to \mu \nu$ \cite{hfag14}.  The lepton universality data by HFAG need to be taken after projecting out a redundant degree in the ratios of the three lepton decay rates \cite{chun1507}.
The precisions of these two data are at the level of 0.1\% and thus sensitive enough to probe L2HDM corrections to the $\tau$ lepton vertices.
Although not significant, we also include all the experimental inputs on  the ten  tau decay (Michel) parameters \cite{pdg14} including all the correlation coefficients \cite{stahl}.

In Section.~\ref{sec:L2HDM} and \ref{sec:gm2}, we briefly summarize the basic properties of L2HDM and its contribution to the muon anomalous magnetic moment relevant for our analysis. 
Section.~\ref{sec:Zdecay} is devoted to the analysis of L2HDM correction to lepton universality in $Z$ decays.
Section.~\ref{sec:taudecay} provides a systematic treatment of the lepton universality data involving the charged current as well as the measurement of the tau decay parameters.
Combining all the leptonic precision data, we show how the parameter space of $(m_A, \tan\beta)$ is constrained in Section.~\ref{sec:result}, and conclude in Section.~\ref{sec:conc}.

\section{Leptophilic 2HDM}
\label{sec:L2HDM}

The most general scalar potential of 2HDM invariant under  the $Z_2$ symmetry, 
$\Phi_{1,2} \to \mp \Phi_{1,2}$, is
\begin{eqnarray} \label{scalar-potential}
	V &=& m_{11}^2 |\Phi_1|^2
	+ m_{22}^2 |\Phi_2|^2 - m_{12}^2 (\Phi_1^\dagger \Phi_2 + \Phi_2^\dagger \Phi_1 ) \nonumber \\
	&& + {\lambda_1\over2} |\Phi_1|^4 + {\lambda_2 \over 2} |\Phi_2|^4
	+ \lambda_3 |\Phi_1|^2 |\Phi_2|^2 + \lambda_4 |\Phi_1^\dagger \Phi_2|^2 + {\lambda_5 \over 2}
	\left[ (\Phi_1^\dagger \Phi_2)^2 + ( \Phi_2^\dagger \Phi_1)^2\right],
\end{eqnarray}
including a (soft) $Z_2$ breaking term $m^2_{12}$.  Upon the electroweak symmetry breaking by the vacuum expectation values (VEVs) $\langle \Phi^0_{1,2} \rangle \equiv v_{1,2}/\sqrt{2}$,  one has five mass eigenstates $h, H,A$ and $H^\pm$ from the two Higgs doublet components: 
$\Phi_{1,2} = \left(\eta^+_{1,2}, {1\over\sqrt{2}} (v_{1,2} + \rho_{1,2} + i \eta^0_{1,2})\right)^{{\rm T}}$
where $v=\sqrt{v_1^2+v_2^2} = 246$ GeV.
Projecting out the longitudinal components of the massive gauge bosons $W^\pm$ and $Z^0$, 
the massive charged and  neutral CP-odd bosons in the limit of negligible CP violation are given by 
\begin{equation}
 H^\pm,A = s_\beta\, \eta_1^{\pm,0}  - c_\beta\, \eta_2^{\pm,0}\,,
\end{equation}
where the angle $\beta$ defines the VEV ratio: $t_\beta\equiv \tan\beta =v_2/v_1$, and $s_\beta \equiv \sin\beta$ and $c_\beta \equiv \cos\beta$. 
Two neutral CP-even bosons are diagonalized by another angle $\alpha$:
\begin{equation}
 h = c_\alpha \,\rho_1 - s_\alpha\, \rho_2, \quad
 H = s_\alpha\, \rho_1 + c_\alpha\, \rho_2\,,
\end{equation}
where $h$ is assumed to be the SM-like Higgs boson with $m_h=125$ GeV.
The gauge couplings of $h$ and $H$ are determined by
\begin{equation}
 {\cal L}_{\rm gauge} = g_V m_V \big(s_{\beta-\alpha} h + c_{\beta-\alpha} H \big) VV\,,
\end{equation}
where $V=W^\pm$ or $Z$ and the SM limit corresponds to
$s_{\beta-\alpha} \to 1$. 

The general Yukawa couplings of 2HDMs are written as
\begin{eqnarray}\label{eqn:genYuk}
-{\cal L}^{\rm 2HDMs}_{\rm Yukawa} &=&
\sum_{f=u,d,l} {m_f\over v}
\left( y_f^h h \bar{f} f + y_f^H H \bar{f} f - i y_f^A A \bar{f} \gamma_5 f \right) \\
&& \nonumber
+\left[ \sqrt{2} V_{ud} H^+  \bar{u} \left( {m_u\over v} y^A_u P_L  + {m_d \over v} y^A_d P_R\right) d
+\sqrt{2} {m_l \over v} y_l^A H^+ \bar{\nu} P_R l + h.c.\right].
\end{eqnarray}
Type-X 2HDM assigns the odd $Z_2$ parity only to the right-handed leptons and thus couples $\Phi_2$ to up/down-type quarks and $\Phi_1$ to charged leptons. As a consequence,  the normalized Yukawa couplings $y^{h,H,A}_f$ are given in the Table~\ref{tab:YukL2HDM}.
\begin{table}[t] \label{Yuk}
\center
\begin{tabular}{|l|ccc|ccc|ccc|}
\hline
& $y_u^A$ & $y_d^A$ & $y_l^A$ & $y_u^H$ & $y_d^H$ & $y_l^H$ & $y_u^h$ & $y_d^h$ & $y_l^h$\\
\hline
~Type X~~ &$\cot\beta$ & $-\cot\beta$ & $\tan\beta$ & $\frac{\sa}{\sb}$ & $\frac{\sa}{\sb}$ & $\frac{\ca}{\cb}$ & $\frac{\ca}{\sb}$ & $\frac{\ca}{\sb}$ & $-\frac{\sa}{\cb}$ \\
\hline 
\end{tabular}
\caption{\label{tab:YukL2HDM}The Yukawa couplings of neutral Higgs bosons in Type X 2HDM.}
\end{table}

Let us recall that the tau Yukawa coupling of the SM Higgs in Type X
can be expressed as
\begin{equation} \label{ytau}
y^h_l = -{s_\alpha \over c_\beta} = s_{\beta-\alpha} - t_\beta c_{\beta-\alpha}.
\end{equation}
Thus, the decoupling/alignment limit of $c_{\beta-\alpha}\to 0$ reproduces not only the usual SM (right-sign) coupling $y^h_l \to +1$ but also the wrong-sign coupling $y^h_l \to  -1$ (with $c_{\beta-\alpha} \approx  2/t_\beta$) compatible with the LHC data \cite{ferreira1410} in the large $t_\beta$ domain of our interest. The other couplings are recasted by
\begin{eqnarray}
&& y^H_l= t_\beta s_{\beta-\alpha} +c_{\beta-\alpha}\to t_\beta, \\ \nonumber
&& y^h_{u,d}=s_{\beta-\alpha}+c_{\beta-\alpha}/t_\beta \to 1, 
\quad y^H_{u,d} = -y^h_l/t_\beta .
\end{eqnarray} 

It is useful to take $\lambda_1$ as a free parameter, and then express the other four couplings $\lambda_{2,3,4,5}$ in terms of  $y^h_l$, $s_{\beta-\alpha}$ and the mass parameters: 
\begin{eqnarray} \label{lambdas}
\lambda_2 v^2 &\approx& s^2_{\beta-\alpha} m_h^2
\,,  \\ \nonumber
\lambda_3 v^2 &\approx& 2 m^2_{H^\pm}
-(s^2_{\beta-\alpha} + s_{\beta-\alpha} y^h_l) m_H^2 + s_{\beta-\alpha} y^h_l m_h^2
\,, \\ \nonumber
\lambda_4 v^2 &\approx&  -2 m^2_{H^\pm} + s^2_{\beta-\alpha} m^2_H + m_A^2 
\,, \\ \nonumber
\lambda_5 v^2 &\approx&  s^2_{\beta-\alpha} m^2_H - m_A^2 \,,\nonumber
\end{eqnarray}
where we have used the relation (\ref{ytau}) neglecting all the terms suppressed by
$1/t_\beta^2$.

As was analyzed in Ref.~\cite{broggio1409}, the explanation of the muon $g-2$ in consistent with 
electroweak precision data  requires  $m_A \ll m_H \approx m_{H^\pm}$.  A custodial symmetry 
is realized to fulfill the electroweak precision test in the limit of a very light pseudoscalar $A$ 
if heavier neutral and charged scalars are almost degenerate \cite{gerard0703}.
One can then find it easy to satisfy the vacuum stability conditions: 
$\lambda_{1,2}>0,~~\lambda_3>-\sqrt{\lambda_1
	\lambda_2},~~|\lambda_5|<\lambda_3+\lambda_4+\sqrt{\lambda_1 \lambda_2}$ within the perturbative limit of $\lambda_1 < 4\pi$. 
In the right-sign (RS) domain of the lepton (tau) Yukawa coupling ($y^h_l s_{\beta-\alpha} \to +1$),
 one finds a strong upper limit of  \cite{broggio1409}
\begin{equation}
 m_A \ll m_{H^\pm} \approx m_H \lesssim 250 \mbox{GeV} \quad \mbox{(RS)}
\end{equation}
where $m_A > m_h/2$ has to be imposed as the $hAA$ coupling, $\lambda_{hAA} v \propto (\lambda_3 +\lambda_4-\lambda_5) v^2 \approx m_h^2 + 2 m_A^2$, is sizeable. 
On the other hand,  in the wrong-sign (WS) domain ($y^h_l s_{\beta-\alpha} \to -1$),
the heavy  boson masses up to the perturbativity limit \cite{wang1412}
\begin{equation}
 m_A \ll m_{H^\pm} \approx m_H \lesssim \sqrt{4\pi} v \quad \mbox{(WS)}
\end{equation}
are allowed and one can also have a much lighter pseudoscalar, $m_A < m_h/2$, 
in the limit of $\lambda_{hAA} v \propto (\lambda_3 +\lambda_4-\lambda_5) v^2 \approx -(s^2_{\beta-\alpha}+s_{\beta-\alpha} y^h_l) m_H^2+ s_{\beta-\alpha} y^h_l m_h^2 + 2 m_A^2 \to 0$.

\section{The $\boldsymbol{(g-2)_{\mu}}$ in L2HDM}
\label{sec:gm2}

For our analysis, we use the value of the muon $g-2$ discrepancy obtained in Ref.~\cite{broggio1409}
considering all the updated SM calculations:
\begin{equation} \label{Damu}
\delta a_\mu \equiv a_\mu^{\rm EXP} - a_\mu^{\rm SM} = + 262\, (85) \times 10^{-11}.
\end{equation}
For the completeness, let us briefly summarize the one- and two-loop contributions to the muon $g-2$ in a 2HDM.

The one-loop contributions to $a_{\mu}$ of the neutral and charged bosons are
\be
	\delta a_\mu^{\mbox{$\scriptscriptstyle{\rm 2HDM}$}}({\rm 1loop}) =
	\frac{G_F \, m_{\mu}^2}{4 \pi^2 \sqrt{2}} \, \sum_j  \left (y_{\mu}^j \right)^2  r_{\mu}^j \, f_j(r_{\mu}^j),
\label{amuoneloop}
\end{equation}
where $j =  \{h, H, A , H^\pm\}$, $y^{A,H,H^\pm} \approx t_\beta$, $|y^h| \approx 1$,
$r_{\mu}^ j =  m_\mu^2/m_j^2$, and
\begin{eqnarray}
	f_{h,H}(r) &=& \int_0^1 \! dx \,  { x^2 ( 2- x) \over 1 - x + r x^2} 
	=- \ln r - 7/6 + O(r), 
\label{oneloopintegrals1} \\ \nonumber
	f_A (r) &=& \int_0^1 \! dx \,  { -x^3  \over 1 - x +r  x^2}
	= +\ln r +11/6 + O(r), 
\label{oneloopintegrals2} \\ \nonumber
	f_{H^\pm} (r) &=& \int_0^1 \! dx \, {-x (1-x)  \over 1 - (1-x) r}
	=-1/6 + O(r),
\label{oneloopintegrals3} \nonumber
\end{eqnarray}
showing that 
$f_{H^\pm}(r)$ is suppressed with respect to $f_{{h,H,A}}(r)$, and 
$h$ and $H$ ($A$ and $H^\pm$) give positive (negative) contributions.

The two-loop Barr-Zee type diagrams with effective
$h\gamma \gamma$, $H\gamma \gamma$ or  $A\gamma \gamma$ vertices generated
by the exchange of heavy fermions give
\be
	\delta a_\mu^{\mbox{$\scriptscriptstyle{\rm 2HDM}$}}({\rm 2loop-BZ}) = \frac{G_F \, m_{\mu}^2}{4 \pi^2 \sqrt{2}} \, \frac{\alpha}{\pi}
	\, \sum_{i,f}  N^c_f  \, Q_f^2  \,  y_{\mu}^i  \, y_{f}^i \,  r_{f}^i \,  g_i(r_{f}^i),
\label{barr-zee}
\end{equation}
where $i = \{h, H, A\}$, $r_{f}^i =  m_f^2/m_i^2$, and $m_f$, $Q_f$ and $N^c_f$ are the mass, electric charge and number of color degrees of freedom of the fermion $f$ in the loop. The functions $g_i(r)$ are
\be
\label{2loop-integrals}
	g_i(r) = \int_0^1 \! dx \, \frac{{\cal N}_i(x)}{x(1-x)-r} \ln \frac{x(1-x)}{r},
\ee
where ${\cal N}_{h,H}(x)= 2x (1-x)-1$ and ${\cal N}_{A}(x)=1$.

Note that the enhancement factor $m_f^2/m_{\mu}^2$ of the two-loop formula \eqref{barr-zee}
can overcome the additional loop suppression factor $\alpha / \pi$, and makes
the two-loop contributions may  become larger than the one-loop ones.
Moreover, contrary to the one-loop contribution, the two-loop functions involving $A$ ($h,H$) are positive (negative), and thus small $m_A$ and large $\tan \beta$ in Type X can generate 
a dominant contribution which can account for the observed $\delta a_{\mu}$ discrepancy
preferring larger $m_H$ to reduce the negative two-loop contribution.  

Recently, more complete computation of the muon $g-2$ coming from additional two-loops involving extra Higgs bosons has been made
\cite{ilisie15}.  We check that their contribution is negligible in the parameter space of our interest and thus
not included in our analysis.
In the following, we will show the 1$\sigma$ and 2$\sigma$ regions in the $m_A$--$t_\beta$ space
(with $m_H=m_{H^\pm}=100,200,300,400$ GeV) considering the additional L2HDM contribution 
$\delta a_{\mu}^{\mbox{$\scriptscriptstyle{\rm 2HDM}$}}  = \delta a_\mu^{\mbox{$\scriptscriptstyle{\rm 2HDM}$}}({\rm 1loop}) + \delta a_\mu^{\mbox{$\scriptscriptstyle{\rm 2HDM}$}}({\rm 2loop-BZ})$  (without the $h$ contribution)
to compare with constraints coming from lepton universality tests.

\section{Lepton universality in Z decays}
\label{sec:Zdecay}

In L2HDM, a sizeable correction to lepton universality can arise from different leptonic Yukawa couplings of  
the extra Higgs bosons, in particular, the tau Yukawa coupling of $m_\tau t_\beta/v$ \eqref{eqn:genYuk}.
The $Z$ boson couplings to fermions and the extra Higgs bosons are
\begin{equation}
-{\cal L} = {g\over c_W} Z^\mu \big\{ \bar{f} \gamma_\mu (g_L P_L + g_R P_R) f
+i (-{1\over2} + s^2_W) H^+ \overleftrightarrow{\partial_\mu}  H^- 
+ A \overleftrightarrow{\partial_\mu} H \big\}\,,
\end{equation}
where $s_W=\sin\theta_W$, and $g_{L,R}= g^0_{L,R} +\delta g_{L,R}$ with
$g^0_{L,R}=T_3(f_{L,R}) -Q(f_{L,R}) s_W^2$, and the small corrections by $c_{\beta-\alpha}\approx0$ 
are neglected in the Higgs sector.  The 2HDM contribution to $Z\to f \bar{f} $ was first calculated in Ref.~\cite{denner91} and recalculated in Ref.~\cite{abe1504}. We rederived it and confirmed that all the results agree with each other in the limit of our interest.

For $f=e,\mu$ and $\tau$ in L2HDM, the one-loop contributions of L2HDM to $\delta g_{L,R}$ 
are given by 
\begin{eqnarray} \label{dgLR}
\delta g^{\rm 2HDM}_L &=& {1\over 16\pi^2} {m_f^2 \over v^2} t_\beta^2 \,
\bigg\{
 -{1\over2} B_Z(r_A)- {1\over2} B_Z(r_H) -2 C_Z(r_A, r_H)
   \nonumber \\
&&  + s_W^2 \left[ B_Z(r_A) + B_Z(r_H) + \tilde C_Z(r_A) + \tilde C_Z(r_H) \right] \bigg\} 
\,,\\ \nonumber
\delta g^{\rm 2HDM}_R &=& {1\over 16\pi^2} {m_f^2 \over v^2} t_\beta^2 \,
\bigg\{ 2 C_Z(r_A, r_H) - 2 C_Z(r_{H^\pm}, r_{H^\pm}) 
+ \tilde C_Z(r_{H^\pm}) - {1\over2} \tilde C_Z(r_A)  - {1\over2} \tilde C_Z(r_H)      \\ \nonumber
&&  +  s_W^2 \left[ B_Z(r_A) + B_Z(r_H) + 2 B_Z(r_{H^\pm})
+\tilde C_Z(r_A) + \tilde C_Z(r_H) + 4C_Z(r_{H^\pm},r_{H^\pm}) \right] \bigg\} 
\,,
\end{eqnarray}
where $r_\phi = m_\phi^2/m_Z^2$ with $\phi=A,H, H^\pm$ and the loop functions are given by
\begin{eqnarray} 
\label{loopftn}
B_Z(r) &=& -{\Delta_\epsilon \over 2} -{1\over4} + {1\over2} \log(r) \,,\\ 
\nonumber
C_Z(r_1,r_2) &=& {\Delta_\epsilon \over4} -{1\over2} \int^1_0 d x \int^x_0 d y\,
\log[ r_2 (1-x) + (r_1 -1) y + x y] \,,\\ 
\nonumber
\tilde C_Z(r) &=& {\Delta_\epsilon \over2}+{1\over2} - r\big[1+\log(r) \big]
+r^2 \big[ \log(r) \log(1+r^{-1}) -{\rm dilog}(-r^{-1}) \big] \\ \nonumber
&& -{i \pi\over2}
\left[ 
	1 - 2r + 2r^{2}\log(1+r^{-1})
\right].
\end{eqnarray}
Here we took the limit $m_\tau \to 0$, and kept the renormalization constant $\Delta_\epsilon=2/\epsilon - \gamma +\log(4\pi)$ in the dimensional regularization with $d=4-\epsilon$, which cancels out in the final 
expressions (\ref{dgLR}). In Appendix.~\ref{apdx:PV}, we provide the explicit one-loop formulae 
in terms of the Passarino-Veltman two- and three-point loop functions \cite{PassarinoVeltman}.
Let us remark that the one-loop corrections (\ref{dgLR}) are suppressed by $1/r$ when
$r=r_{A,H,H^\pm} \gg 1$ as expected in the decoupling limit. However, we have  $r_A< 1 <  r_{H}\approx r_{H^\pm}$ in our favorable parameter space, and the loop corrections become larger 
for higher hierarchy,  $r_A \ll r_{H, H^\pm}$.

\medskip

Precision electroweak measurements were performed by the SLD and LEP experiments with data taken at the Z resonance which provides also lepton universality test in Z decay through the ratios of the leptonic 
branching fractions \cite{lepew05}:
\begin{eqnarray} \label{lu-zdecay}
{\Gamma_{Z\to \mu^+ \mu^-}\over \Gamma_{Z\to e^+ e^- }} &=& 1.0009 \pm 0.0028
\,,\\ \nonumber
{\Gamma_{Z\to \tau^+ \tau^- }\over \Gamma_{Z\to e^+ e^- }} &=& 1.0019 \pm 0.0032
\,,
\end{eqnarray}
with a correlation of $+0.63$.  From this, we calculate $\delta_{ll} = (\Gamma_{Z\to l^+ l^-}/\Gamma_{Z\to e^+ e^- })-1$ for $l=\mu$ and $\tau$ to compare with 
the L2HDM correction:
\begin{eqnarray}
 \delta_{\mu\mu} &\simeq& 0
 \,, \\ \nonumber
  \delta_{\tau\tau} &=& {2 g_L^e{\rm Re}(\delta g^{\rm 2HDM}_L)+ 2 g_R^e{\rm Re}(\delta g^{\rm 2HDM}_R) \over {g_L^e}^2 + {g_R^e}^2 }\,,
\end{eqnarray}
where we use the SM value $g_L^e=-0.27$ and $g_R^e=s_W^2=0.23$ which have also been measured
by the electroweak precision test \cite{lepew05}.

\section{Lepton universality and tau decays}
\label{sec:taudecay}

In the  large $\tan\beta$ limit of Type II and X 2HDM, there appear two important corrections to $\tau$ 
decays: tree-level contribution of the charged boson and one-loop corrections from the extra bosons which
have been analyzed extensively in Ref.~\cite{maria04}.
A recent study \cite{abe1504} showed that it can constrain strongly the L2HDM parameter space
in favor of the muon $g-2$.  We reconfirm the results leading to the following tree and loop corrections, e.g.,
to the $\tau$ decay rate $\Gamma_{\tau \to l \nu \nu} =\Gamma_{\tau \to l \nu \nu}^{\rm SM} (1+2 \delta_{\rm tree} + 2 \delta_{\rm loop})$:
\begin{eqnarray} \label{deltas}
\delta_{\rm tree} &=& {m_\tau^2 m_\mu^2 \over 8 m^4_{H^\pm}} t^4_\beta
- {m_\mu^2 \over m^2_{H^\pm}} t^2_\beta {g(m_\mu^2/m^2_\tau) \over f(m_\mu^2/m_\tau^2)}, \\
\delta_{\rm loop} &=& {1 \over 16 \pi^2} { m_\tau^2 \over v^2}  t^2_\beta
\left[1 + {1\over4} \left( H(x_A) + s^2_{\beta-\alpha} H(x_H) + c^2_{\beta-\alpha} H(x_h)\right)
\right]\,, \nonumber
\end{eqnarray}
where $f(x)\equiv 1-8x+8x^3-x^4-12x^2 \ln(x)$, $g(x)\equiv 1+9x-9x^2-x^3+6x(1+x)\ln(x)$ and
$H(x_\phi) \equiv \ln(x_\phi) (1+x_\phi)/(1-x_\phi)$ with $x_\phi=m_\phi^2/m_{H^{\pm}}^2$.

In practice, one can neglect the contribution from the SM scalar $h$ which is proportional to a small number 
$c_{\beta-\alpha}^2$ as was the case in the previous section.
It is now worth noticing that the one-loop correction in Eq.~\eqref{deltas} shows a non-decoupling behavior, 
that is, it is not suppressed by the large mass and remains constant as far as the mass ratios are kept. 
Furthermore, it vanishes in the limit of $m_A=m_H=m_{H^\pm}$. As a big hierarchy 
$m_A \ll m_H \approx m_{H^\pm}$ (and also large $\tan\beta$) is required, 
one can expect a large correction to tau decays.

\medskip

In the previous analyses \cite{abe1504,chun1507}, only partial data set from the lepton universality test 
by HFAG \cite{hfag14} was taken. In this work, we take the full data set properly including all the correlation effect.
HFAG provided three ratios of couplings from pure leptonic processes, $l' \to l \nu \nu$, and two ratios
from semi-hadronic processes, $\tau \to \pi/K \nu$ and $\pi/K \to \mu \nu$: 
\begin{eqnarray} \label{hfag-data}
&&
\left( g_\tau \over g_\mu \right) = 1.0011 \pm 0.0015, \quad
\left( g_\tau \over g_e \right) = 1.0029 \pm 0.0015, \quad
\left( g_\mu \over g_e \right) = 1.0018 \pm 0.0014, 
\nonumber\\
&&
\left( g_\tau \over g_\mu \right)_\pi = 0.9963 \pm 0.0027, \quad
\left( g_\tau \over g_\mu \right)_K= 0.9858 \pm 0.0071,
\end{eqnarray}
with the correlation matrix for the above five observables:
\begin{equation} \label{hfag-corr}
\left(
\begin{array}{ccccc}
1 & +0.53 & -0.49 & +0.24 & +0.12 \\
+0.53  & 1     &  + 0.48 & +0.26    & +0.10 \\
-0.49  & +0.48  & 1       &   +0.02 & -0.02 \\
+0.24  & +0.26  & +0.02  &     1    &     +0.05 \\
+0.12  & +0.10  & -0.02  &  +0.05  &   1 
\end{array} \right) .
\end{equation}
The quantities in Eq.~\eqref{hfag-data} can be calculated in L2HDM as follows:
\begin{eqnarray} \label{deltas-data}
&&
\left( g_\tau \over g_\mu \right) = 1+ \delta_{\rm loop}, \quad
\left( g_\tau \over g_e \right) =1+ \delta_{\rm tree}+ \delta_{\rm loop}, \quad
\left( g_\mu \over g_e \right) = 1+ \delta_{\rm tree}, 
\nonumber\\
&&
\left( g_\tau \over g_\mu \right)_\pi = 1+ \delta_{\rm loop}, \quad
\left( g_\tau \over g_\mu \right)_K= 1+ \delta_{\rm loop}.
\end{eqnarray}
As is obvious from the definition, one can see the relation $(g_\tau/g_\mu)(g_\mu/g_e)/(g_\tau/g_e)$
which shows three pure leptonic data are not independent. Due to this, one can  find that the covariance matrix constructed from the data \eqref{hfag-data}--\eqref{hfag-corr} has a vanishing eigenvalue, and thus the 
corresponding degree has to be removed to form a proper chi-squared variable. 

Notice that the best-fit values of the semi-hadronic data are in the negative side, which are opposite to  
those of the pure leptonic data, and that the one-loop correction  $\delta_{\rm loop}$ is always negative.
As we will see, the proper treatment of the lepton universality data in the charged current gives much weaker bounds than the previous ones \cite{abe1504,chun1507} which used a partial data set.

\medskip

The Lorentz structure (spin and chirality) of the charged current is determined experimentally
by measuring the $\tau$ decay (Michel) parameters \cite{pdg14}. Although its precision is at the level of 
1 \%, it can play an important role to break down the degeneracy in the lepton universality constraint on the $\tau$ decay \cite{abe1504}.  Adding the L2HDM contribution to the SM prediction for the $\tau$ decay parameters in the processes 
$\tau \to l \nu \nu$ (with $l=e,\mu$), one finds
\begin{eqnarray} \label{michel}
&&
\rho(e)= {3\over4},\quad 
\xi(e)=1, \quad
\delta\xi(e)={3\over4},
\quad
\rho(\mu)= {3\over4},
\\ \nonumber
&&
\eta(\mu)= - {2 z (1 + \delta_{\rm loop}) \over 4 + z^2},\quad 
\xi(\mu) = {4(1+\delta_{\rm loop})^2-z^2 \over 4 (1+\delta_{\rm loop})^2+ z^2}, \quad  
\delta\xi(\mu)={3\over4} {4(1+\delta_{\rm loop})^2-z^2 \over 4 (1+\delta_{\rm loop})^2+ z^2}, \quad  
\\ \nonumber
&&
\xi(\pi)=-1, \quad
\xi(\rho)=-1, \quad
\xi(a_1)=-1,
\end{eqnarray}
where the tree-level contribution from $z=m_\mu m_\tau t_\beta^2/m_{H^\pm}^2$ and 
the one-loop contribution from $\delta_{\rm loop}$ are included for completeness 
although the loop correction is too small to give a sizeable contribution.
The PDG determination of the above 10 parameters (without assuming lepton universality) is 
summarized in Appendix.~\ref{apdx:taudecayparam}, and will be combined with the HFAG data in our final result (Fig.~\ref{fig:loops}). 
The inclusion of the $\tau$ decay parameters slightly more constrains the parameter space.

\section{Results}
\label{sec:result}

\begin{figure}[tbp]
\centering
\includegraphics[width=0.4\textwidth]{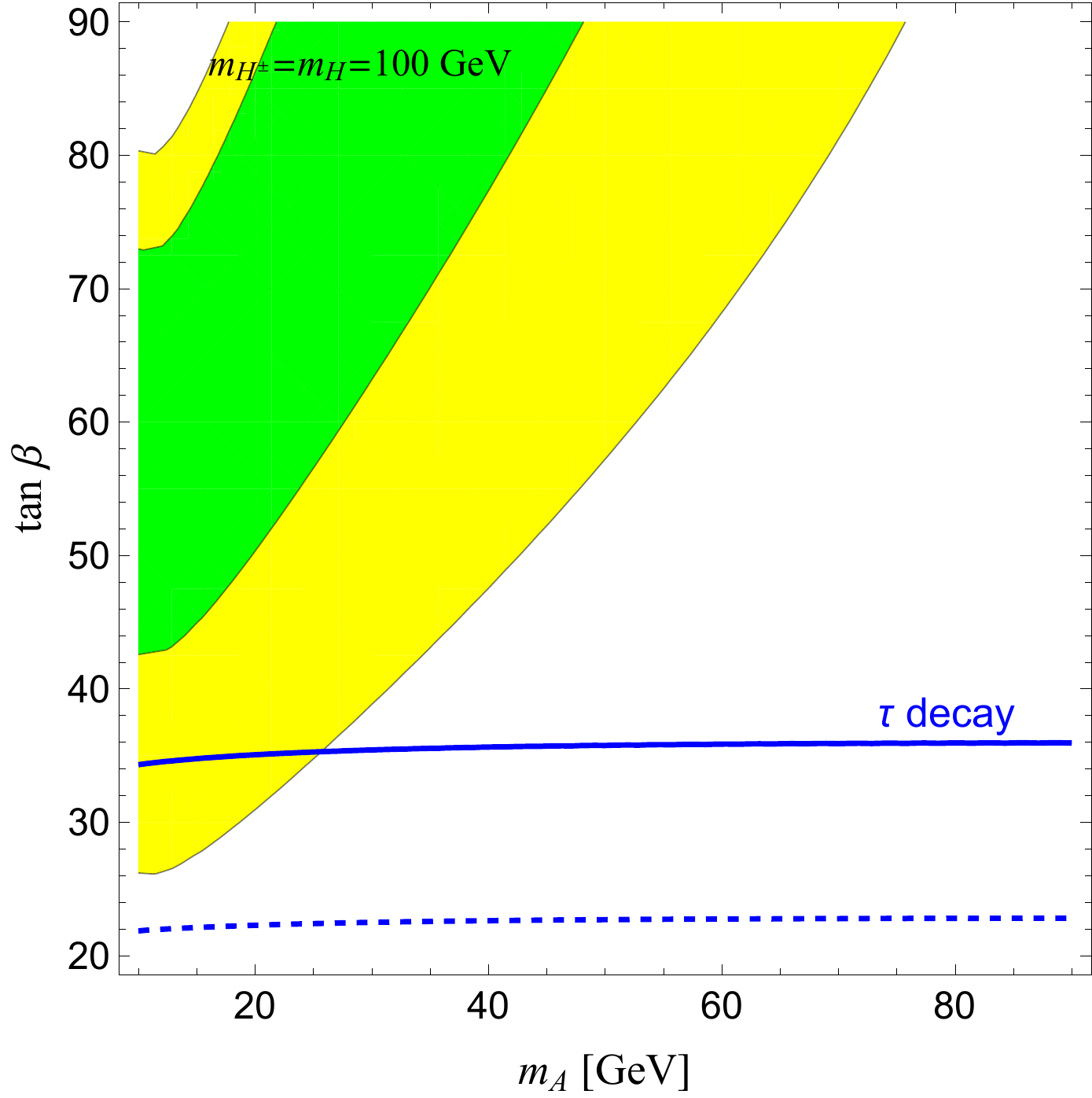}
\quad
\includegraphics[width=0.4\textwidth]{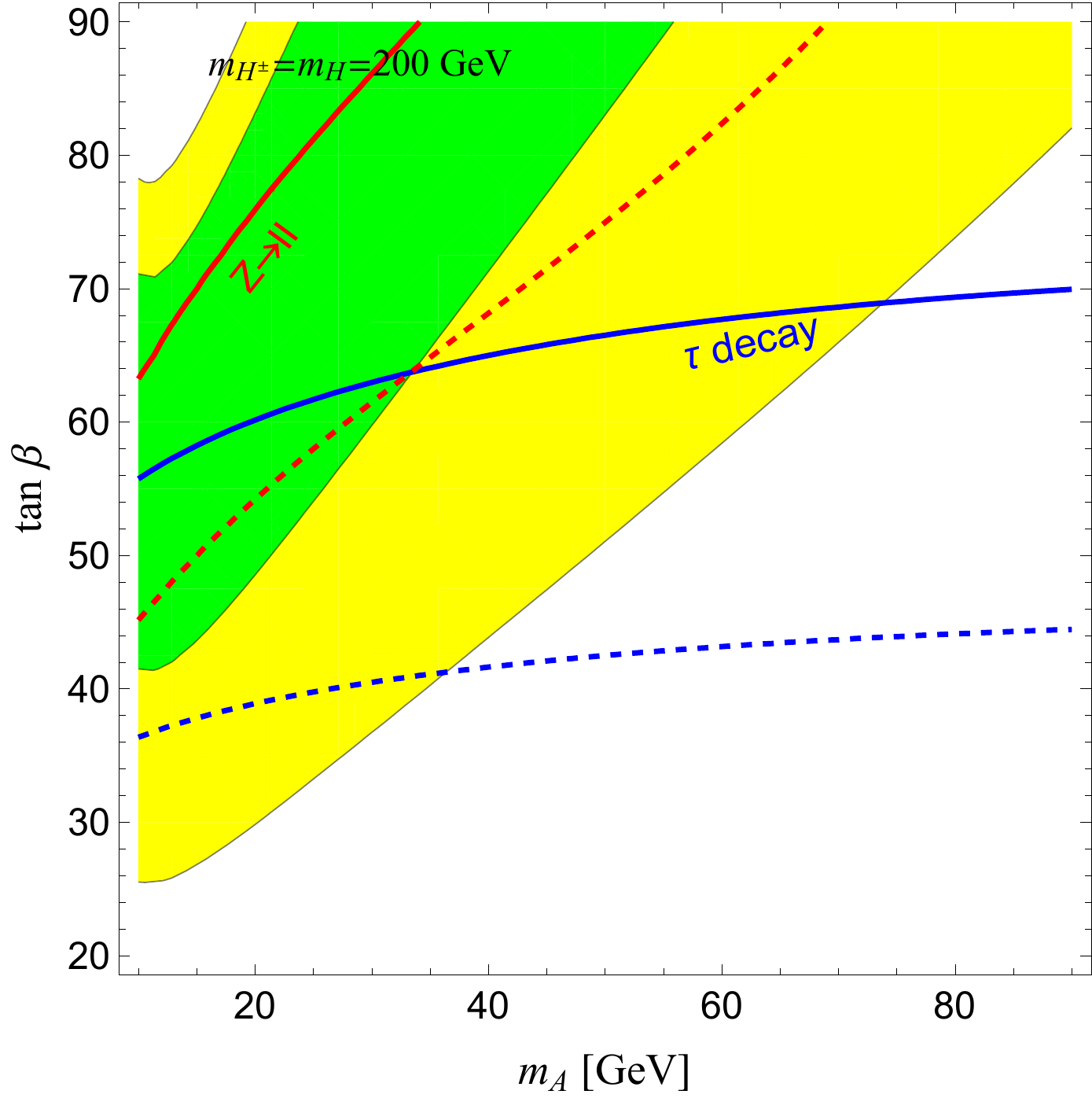}
\vfill
\includegraphics[width=0.4\textwidth]{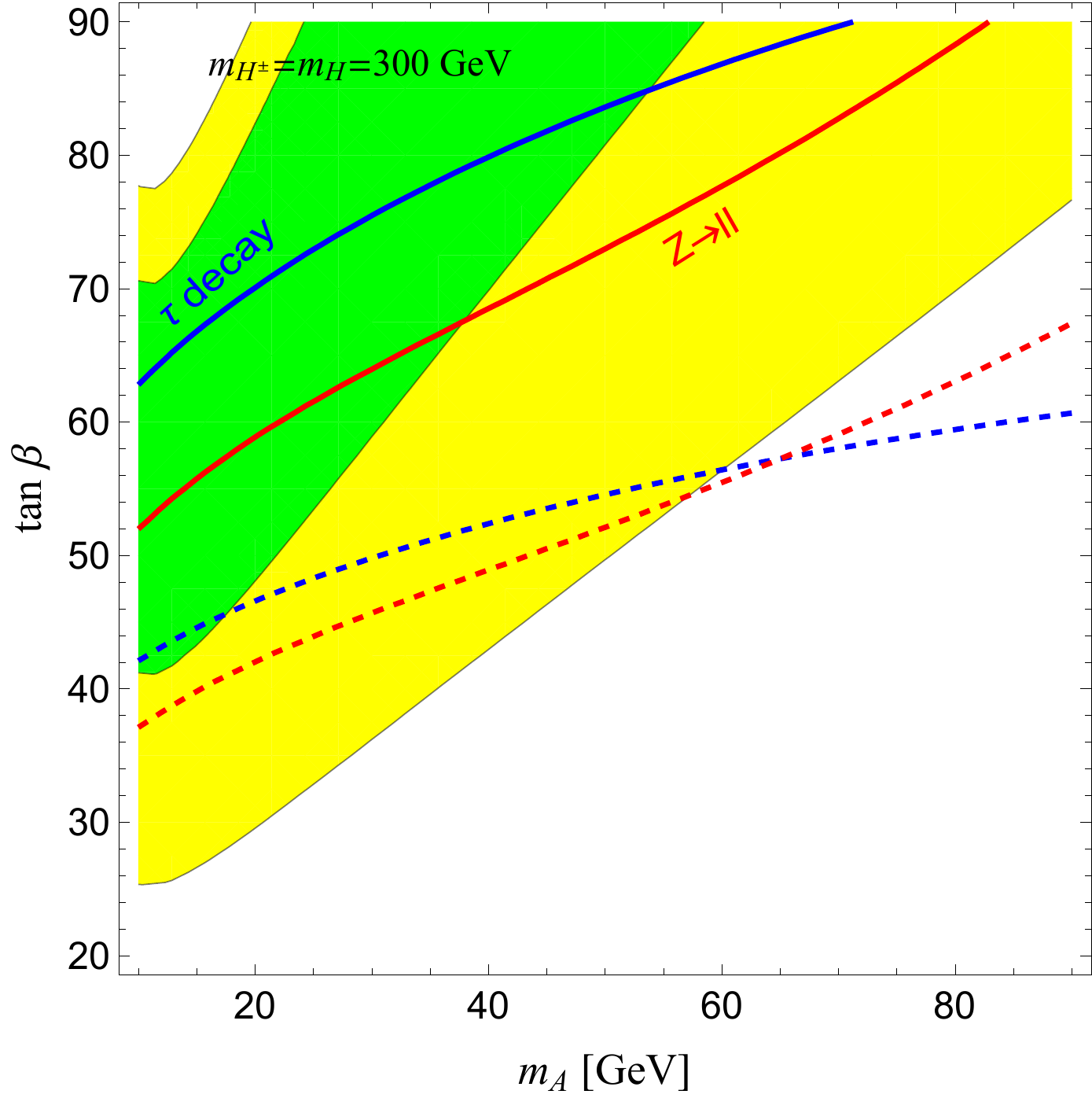}
\quad
\includegraphics[width=0.4\textwidth]{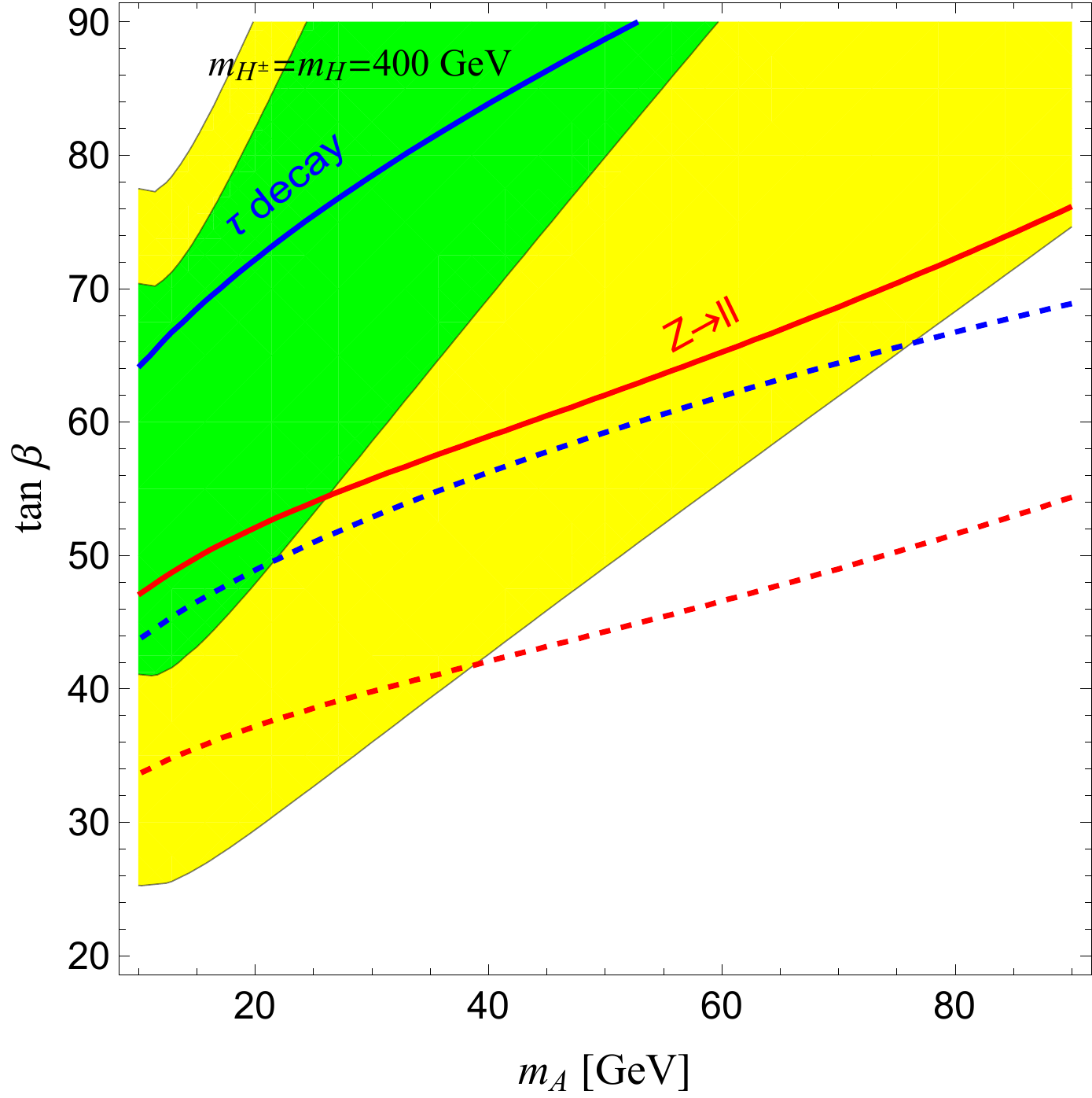}
\caption{\label{fig:loops}Allowed at $1\sigma$ and $2\sigma$ CL are  the regions inside the green (inner) and yellow (outer) shaded areas by the muon $g-2$; below the red dashed (lower) and red solid (upper) lines by the lepton universality test in $Z$ decays; below the blue dashed (lower) and blue solid (upper) lines by the lepton universality test with  $\tau$ decays, respectively. 
}
\end{figure}

Fig.~\ref{fig:loops} summarizes the results of our analyses on the allowed regions at 1$\sigma$ and 2$\sigma$ 
for the muon $g-2$ (colored/shaded region), lepton universality in $Z$ decays (red lines), and 
lepton universality with $\tau$ decays (blue lines) in the plane of ($m_A, \tan\beta$). 
Four different values of $m_{H}=m_{H^\pm}=100,200,300$ and $400$ GeV were chosen in a way that 
the precision electroweak test is fulfilled by the degeneracy of the two heavier Higgs boson masses.

First, notice that the allowed region for the muon $g-2$ gets larger for heavier $H/H^\pm$ as it reduces 
the negative contribution from $H$.  Let us recall that we used the deviation of the muon $g-2$ given in Eq.~\eqref{Damu}, allowing a larger area than in Ref.~\cite{abe1504}.

The constraints from the lepton universality in $Z$ decays become 
stricter for larger $m_H=m_{H^\pm}$ due to larger hierarchy $m_A \ll m_{H, H^\pm}$. 
After properly including both data for $\delta_{\tau\tau}$ and $\delta_{\mu\mu}$ with the corresponding correlation coefficient, we found that our bounds are similar to those in Ref.~\cite{abe1504}.
On the other hand, the constraints from the lepton universality in $\tau$ decays are stronger for smaller 
$m_{H^\pm}$ due to the enhanced tree-level contribution which is independent of $m_A$. 
For heavier $H/H^\pm$, the one-loop correction become more important and the region of 
smaller $m_A$ is much more constrained again due to the large mass splitting $m_A \ll m_{H, H^\pm}$.
Recall that we included the full HFAG data from the pure leptonic and semi-hadronic decays and obtained 
much weaker bounds than those in Refs.~\cite{abe1504,chun1507}. In fact, the lepton universality data with tau decays
show a $2\sigma$ deviation from the SM prediction and thus we obtain a large chi-square minimum, $\chi^2_{min}=15.6$, even for the L2HDM fit. In Fig.~\ref{fig:loops}, we show all the $1\sigma$ and $2\sigma$ regions of $\Delta \chi^2 =  \chi^2-\chi^2_{min}$.

Note that the lepton universality test in tau decays 
provides much stronger bounds for lower $m_{H/H^\pm}$, but the lepton universality test in $Z$ decays becomes more constraining for larger $m_{H/H^\pm}$. 
Even after combining both lepton universality tests,  
there survives still a large region of $(m_A, \tan\beta)$ accommodating the muon $g-2$ deviation at $2\sigma$
for intermediate values of $m_{H/H^\pm}\sim 200-400$ GeV.

\section{Conclusion}
\label{sec:conc}

In the context of 2HDM, the type X (lepton-specific) model at large $\tan\beta$ (L2HDM) can accommodate
the observed deviation of the muon anomalous magnetic moment if the extra Higgs bosons follow the mass spectrum $m_A \ll m_H \approx m_{H^\pm}$.  While no serious constraints can be applied by
hadronic observables, it can be strongly constrained by precision leptonic observables. 
We showed how the consistent combination of the lepton universality data in $Z$ and $\tau$ decays constrains the L2HDM parameter space. Indeed, a large region of  $(m_A, \tan\beta)$ explaining the muon $g-2$ anomaly 
is excluded so that no region is allowed at $1\sigma$.  Still, a sizeable parameter space is viable at $2\sigma$,
particularly, for intermediate values of $m_{H}=m_{H^\pm}$ around  $ 200 \sim 400$ GeV.  

The future improvement of the precision of lepton universality will be crucial to limit further the L2HDM 
parameter space in favor of the muon $g-2$ explanation. Such a light pseudoscalar $A$ could be searched for 
by observing the processes of $AH/H^\pm \to AA+X$ and $h \to AA$ leading to the final states 
of $4\tau$ \cite{chun1507} and $2\mu2\tau$ \cite{atlas1505} in future collider experiments.

\begin{appendix}
\section{Two- and Three-point functions in Z decays}
\label{apdx:PV}

Defining $D_i=(k+ \sum_{k=1}^{i-1} p_k)^2-m_i^2$ with introducing a redundant momentum $p_{i_{max}}$
satisfying the energy momentum conservation,  $\sum_{k=1}^{i_{max}} p_k =0$, 
one gets the following two- and three-point functions
relevant for our calculation \cite{PassarinoVeltman}:
\begin{eqnarray}
B^\mu(p_1^2,p_2^2; m_1^2,m_2^2) 
&=& {(2\pi\mu)^{4-d} \over i \pi^{2}} \int d^d k {k^\mu \over D_1 D_2}  \\ \nonumber
&=& p_1^\mu B_1(p_1^2, p_2^2;  m_1^2, m_2^2)  \,,\\ 
C^\mu(p_1^2,p_2^2,p_3^2; m_1^2, m_2^2, m_3^2) 
&=&   {(2\pi\mu)^{4-d} \over i \pi^{2}} \int d^d k {k^\mu \over D_1 D_2 D_3} \\ \nonumber
&=& p_1^\mu C_1(p_1^2,p_2^2,p_3^2; m_1^2, m_2^2, m_3^2)  \\ \nonumber
&&+ p_2^\mu C_2(p_1^2,p_2^2,p_3^2; m_1^2, m_2^2, m_3^2) \,,\\ 
C^{\mu\nu}(p_1^2,p_2^2,p_3^2; m_1^2, m_2^2, m_3^2) 
&=&   {(2\pi\mu)^{4-d} \over i \pi^{2}} \int d^d k {k^\mu k^\nu \over D_1 D_2 D_3} \\ \nonumber
&=& g^{\mu\nu} C_{00}(p_1^2,p_2^2,p_3^2; m_1^2, m_2^2, m_3^2) \\ \nonumber 
&&+ p_1^\mu p_1^\nu  C_{11} (p_1^2,p_2^2,p_3^2; m_1^2, m_2^2, m_3^2)  \\ \nonumber 
&&+ p_2^\mu p_2^\nu  C_{22} (p_1^2,p_2^2,p_3^2; m_1^2, m_2^2, m_3^2)  \\ \nonumber 
&&+ (p_1^\mu p_2^\nu +p_1^\nu p_2^\mu) C_{12} (p_1^2,p_2^2,p_3^2; m_1^2, m_2^2, m_3^2) \,,
\end{eqnarray}
where $d=4-\epsilon$ is the dimensional regularization parameter and $\mu$ is the renormalization scale.
The loop-functions in Eq.~\eqref{loopftn} are obtained by
\begin{equation}
B_Z\left({m^2 \over m_Z^2}\right) = B_1(0,0;0,m^2) \,,
\end{equation}
\begin{equation}
C_Z\left({m_1^2 \over m_Z^2},{m_2^2 \over m_Z^2}\right) 
= C_{00}(0,0,m_Z^2;m_1^2,0,m_2^2)= C_{00}(0,0,m_Z^2;m_2^2,0,m_1^2)\,,
\end{equation}
\begin{equation}
\tilde C_Z \left({m^2 \over m_Z^2}\right) 
= \big[(d-2) C_{00} + m_Z^2 (C_{12} + C_2)\big](0,0,m_Z^2; 0, m^2, 0) \,,
\end{equation}
taking the renomalization scale $\mu=m_Z$.
Note that $B_1$ and $C_{00}$ contain the renormalization constant $-\Delta_\epsilon/2$ and 
$\Delta_\epsilon/4$, respectively.

\section{Tau decay parameters}
\label{apdx:taudecayparam}

The PDG determination of the $\tau$ decay parameters \cite{stahl} is given by
\begin{eqnarray}
&&
\rho(e)= 0.7475 \pm 0.0097, \quad
\xi(e) = 0.9939 \pm  0.0404, \quad
\delta\xi(e)= 0.7337 \pm  0.0282, \\ \nonumber
&&
\rho(\mu) = 0.7630 \pm  0.0196, \quad
\eta(\mu) = 0.09678 \pm  0.07265, \quad  \\ \nonumber
&&
\xi(\mu) = 1.0288 \pm 0.0589, \quad
\delta\xi(\mu) = 0.7774 \pm 0.0374,   \\ \nonumber
&&
\xi(\pi) = -0.9946 \pm 0.0219, \quad
\xi(\rho) = -0.9941 \pm 0.0084,  \quad
\xi(a_1) = -1.00037 \pm   0.02731,
\end{eqnarray}
with the correlation matrix:
\begin{equation}
\left(
\begin{array}{ccccccccccc}
1 & -0.055 & 0.016 & 0.051 & -0.031 & -0.009 & 0.00052 & -0.039 & 0.019 & -0.00055 \\
-0.055 & 1 & 0.17 & 0.016 & 0.0028 & -0.031 & -0.047 & 0.04 & -0.12 & 0.01 \\ 
0.016 & 0.17 & 1 & 0.0091 & 0.012 &-0.052 & -0.014 &  0.068 & -0.15 & 0.0038 \\
0.0051 & 0.016 & 0.0091 & 1 & 0.7 & 0.12 &  0.017 & -0.042 &-0.0082 &0.0033 \\
-0.031 &0.0028 & 0.012 & 0.7 & 1 & 0.3 & 0.091 & 0.073 & -0.063 & -0.03 \\
-0.009 &-0.031 & -0.052 & 0.12 & 0.3 &1 & 0.015 &  0.027 & -0.063 & 0.013 \\
0.00052 &-0.047 &-0.014 & 0.017 & 0.091 & 0.015 & 1 & 0.06 & -0.087 & 0.0094 \\
 -0.039 &0.04 & 0.068 & -0.042 & 0.073 & 0.027 & 0.06 & 1 & -0.16 & -0.20 \\
0.019 &-0.12 & -0.15 & -0.0082 & -0.063 & -0.063 & -0.087 &-0.16 & 1 & -0.047 \\
-0.00055 & 0.01 & 0.0038 & 0.0033 & -0.03 & 0.013 & 0.0094 & -0.20 & -0.047 & 1 
\end{array}
\right) .
\end{equation}

\end{appendix}

\medskip

{\bf Acknowledgment}: EJC thanks Alberto Lusiani and Achim Stahl for  
providing detailed information on lepton universality tests in $\tau$ and $Z$ decays, 
and  critical discussions on the data analysis.


\end{document}